\documentclass[10p]{elsart}
\usepackage{graphicx}
\usepackage{amsmath}

\begin{document}
\begin{frontmatter}

\title{Slow Kinetics of Capillary Condensation in Confined Geometry: Experiment and Theory}
\author[dpm,ens]{F. Restagno}
\author[ens]{L. Bocquet}
\author[ens]{J. Crassous}
\author[dpm]{E. Charlaix}
\address[dpm]{D\'epartement de physique des mat\'eriaux, 6 rue Amp\`ere, 69622 Villeurbanne cedex (France)}
\address[ens]{Laboratoire de Physique, ENS Lyon, 46 all\'ee d'Italie, 69364 Lyon cedex 07 (France)}

\begin{abstract}
When two solid surfaces are brought in contact, water vapor
present in the ambient air may condense in the region of the
contact to form a liquid bridge connecting the two surfaces : this
is the so-called capillary condensation. This phenomenon has
drastic consequences on the contact between solids, modifying the
macroscopic adhesion and friction properties. In this paper, we
present a survey of the work we have performed both experimentally
and theoretically to understand the microscopic foundations of the
kinetics of capillary condensation. From the theoretical point of
view, we have computed the free energy barrier associated with the
condensation of the liquid from the gas in a confined system.
These calculations allow to understand the existence of very large
hysteresis, which is often associated with capillary condensation.
This results are compatible with  experimental results obtained
with a surface forces apparatus in a vapor atmosphere, showing a
large hysteris of the surface energy of two parallel planes as a
function of their distance. In the second part, we present some
experiments on the influence of humidity on the avalanche angle of
granular media. We show that the ageing in time of this avalanche
angle can be explained by the slow kinetics of capillary
condensation in a random confined geometry.

\end{abstract}

\end{frontmatter}

\newpage
\section{Introduction}
Molecules confined in narrow pores, with pore widths of a few
molecular diameters, can exhibit a wide range of physical
behavior. The introduction of wall forces, and the competition
between fluid-wall and fluid-fluid forces, can lead to interesting
surface driven phase changes, since for a small confinement the
surface effects can be more important than the bulk effects
\cite{Gelb99}. Such effects can be observed in porous materials
which have a large specific area. Porous materials are involved in
many physical, chemical or biological
 processes. Their adsorption properties
 are known to present a variety of behavior related to the texture of the
 porous matrix, which provides an experimental way to analyze the pore size
 distribution. Interpretation of adsorption isotherms in these materials commonly
 involves a well known phenomenon, capillary condensation \cite{Gelb99,Israel,Evans86},
 which corresponds to the condensation of liquid bridges in the pores.
 More fundamentally, capillary condensation is a gas-liquid phase
 transition shifted by confinement. A basic model of confinement is
 provided by the slab geometry, for which the fluid is confined
 between two parallel planar solid walls. The classical theory of capillarity \cite{Evans86}
  predicts that in this geometry the liquid phase condenses when the substrate-liquid surface
 tension $\gamma_{SL}$ is smaller than the substrate-vapor surface tension
  $\gamma_{SV}$,
and when the distance between the surfaces is lower than $H_c$
satisfying the
 Kelvin equation:
\begin{equation}\Delta\rho~\Delta\mu \simeq
{{2(\gamma_{SV}-\gamma_{SL})}/H_c}\end{equation} Here,
 $\Delta\rho=\rho_L-\rho_V$ is the difference between the bulk
 densities of the liquid and the gas phase, $\Delta\mu=\mu_{sat}-\mu$ is
 the (positive) undersaturation and $\mu_{sat}$ is
 the chemical potential at bulk coexistence. If the vapor can be considered as an ideal gas, we have:
 $\Delta\mu=k_BT\ln(p_{sat}/p_{vap})$, where $k_B$ is the Boltzmann's constant, $T$ is the absolute temperature and $p_{sat}/p_{vap}$ the saturated vapor pressure divided by the partial pressure of the vapor.
  Although the equilibrium properties
 of this transition have motivated many experimental \cite{Fisher79,Fisher81a,Christenson88,Crassous94}
 and theoretical studies \cite{Evans85,Evans86,Derjaguin92}, capillary condensation
 presents remarkable dynamical features which are still to be explained.
 The most striking feature is the huge metastability of the coexisting
 phases, which contrasts with the bulk liquid-vapor transition.

 Since capillary condensation is a first order phase
 transition, one should be able to identify a critical nucleus and a
 corresponding free energy barrier away from the spinodal. For sufficiently small
 $H$, it can be shown that the liquid films coating the
solid surfaces become unstable due to fluid-fluid interactions and
grow to fill the slab. This has been carefully studied by several
authors \cite{Crassous94,Christenson94,Forcada93a}).
 In this article we show that, as in the homogeneous nucleation case,
 the shape of the critical nucleus results from the balance between surface
 and volume contributions. The height of the activation barrier is
 so large that it can induce a large metastability of the vapor
 phase. This first theoretical result is compared to
 experiments on capillary condensation in a surface forces
 apparatus. In the second part of this article, we will discuss
 the influence of this slow kinetics of capillary condensation on the mechanical properties
 of a granular material in a humid atmosphere.

\section{Homogeneous nucleation of a liquid phase between ideal surfaces}
\subsection{Method}
Since capillary condensation occurs only in a confined geometry,
the problem which arises in computing an energy barrier for the
vapor/liquid transition is the validity of the macroscopic
concepts of the classical theory of capillarity.

To address this problem we use the following approach: \\i-- We
use a Density Functional Theory (DFT) model for the fluid phase,
taking into account the long range interactions with the solid
surfaces, and study the time evolution of a metastable confined
vapor with a Langevin equation. We perform this study in a two
dimensional geometry and determine the energy barrier associated
with the vapor/liquid transition as a function of the distance
between the walls.\\
ii-- We compare these results with the prediction of the classical
theory of capillarity in two dimensions. The classical theory
gives the correct qualitative behavior and dependency of the
energy barrier as a function of the confinement up to a numerical
prefactor.\\
iii-- We then use the classical theory to calculate the energy
barrier to condensation between two parallel plates in the three
dimensional case, and discuss the order of magnitude obtained.

\subsection{Theoretical calculation of the activation energy in a 2D geometry}

We first use a mesoscopic
 Landau-Ginzburg model for the grand potential of the 2D system confined between
 two walls. In terms of the local density $\rho({r})$, we write the ``excess''
 part of the grand potential $\Omega^{ex}=\Omega+p_{sat}V$, where $p_{sat}$ is
 the pressure of the system at coexistence, as:
 \begin{equation}
 \Omega^{ex}= \int d{r}\left\{ {m\over 2}  \vert\nabla \rho
 \vert^2 + W(\rho) +  \left(\Delta\mu +v_{ext}(z)\right) \rho \right\}
 \label{eq5}
 \end{equation}
 In this equation, $m$ is a phenomenological parameter allowing a simple treatment of inhomogeneous fluids; $v_{ext}(z)$ is the confining external
 potential,
 which we took for each wall as $v_{ext}(z)=-\epsilon (\sigma/(\Delta
 z+\sigma))^3$, with $\Delta z$ the distance to the corresponding wall;
 $\epsilon$ and $\sigma$ have the dimensions of an energy and a distance, $W(\rho)$ can be interpreted as the negative of the
 ``excess'' pressure $\mu_{sat}\rho-f(\rho)-p_{sat}$, with $f(\rho)$ the bulk free-energy
 density \cite{Rowlinsom}. To allow a phase transition in this system, we assume a phenomenological double well
 form for $W(\rho)$ : $W(\rho)=a(\rho-\rho_V)^2(\rho-\rho_L)^2$, where $a$ is a
 phenomenological parameter \cite{Safran}. The system
 is then driven by a non-conserved Langevin equation for $\rho$:
 \begin{equation}
 {\partial\rho \over{\partial t}}= -\Gamma {\delta \Omega^{ex} \over {\delta\rho}} + \eta({r},t)
 \label{langevin}
 \end{equation}
 where $\Gamma$ is a phenomenological friction coefficient and $\eta$ is
 a Gaussian noise field related to $\Gamma$ through the fluctuation-dissipation
 relationship \cite{Chaikin}. This time dependent Landau-Ginzburg model provides a good
 phenomenological description of the dynamics of the density field
 $\rho(\vec{r},t)$ as soon as the dynamics of the nucleation is
 not limited by the crossing of the activation barrier. This model
 has been previously successfully applied to homogeneous
 nucleation \cite{Valls90}.
\begin{figure}
\centering
\includegraphics[width=7cm]{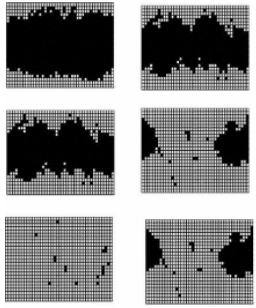}
\caption{Snapshots of the density $\rho(\vec{r})$ of the fluid for
different times $t=2,6,12,18,20,28$. If a point appears in white
it means that the local density at this point is higher than
$0.8$. To obtain $\bar{\rho}(t)$, we make an average on the mean
density in the pore on different realizations of the noise.
}\label{fig10}
\end{figure}
 The units of energy, length
 are such that $\sigma=\epsilon =1$. Time is in units of $t_0=(\Gamma \epsilon\sigma^2)^{-1}$
 with $\Gamma={1\over 3}$. In these units,
 we took $m=1.66$, $a=3.33$, $\rho_L=1$, $\rho_V=0.1$. Typical values of the chemical
 potential and temperature are $\Delta\mu\sim 0.016$, $T\sim 0.06$ (which is
 roughly half the critical temperature in this model). Periodic
 boundary conditions with periodicity
 $L_x$ were applied in the lateral direction.
 The simulated system is initially a gas state filling the whole pore, and its
 evolution is described by equation (\ref{langevin}). A typical evolution (see figure \ref{fig10}) of the
 mean density
 in the slit shows that: i) As expected \cite{Evans85}, due to the long range nature of the external
 potential a thick liquid film of thickness $\ell$
 rapidly forms on both walls on a short
 time scale $\tau_1$ ($\ell\simeq 3.8 \sigma$ and $\tau_1\approx 5 t_0$ in our case)
 ii) Fluctuations of the interfaces around their mean value $\ell$ induce after
 a while a sudden  coalescence of the films. This second process has a characteristic time $\tau$.

\begin{figure}[htbp]
\centering
\includegraphics[width=6cm]{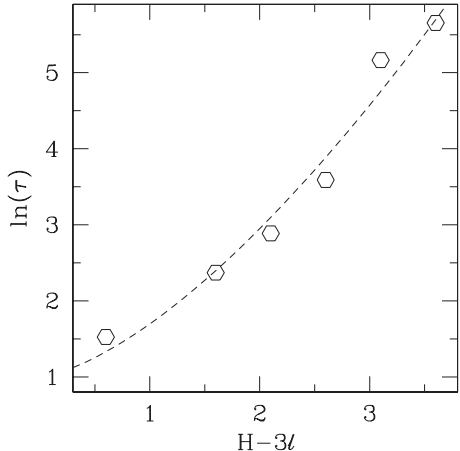}
\caption{Logarithm of condensation time as a function of the
``effective'' width of the slab $H-3\ell$ for fixed
$\Delta\mu=0.016$. The dashed line is the theoretical prediction
$\ln(\tau)=\ln(\tau_0)+\alpha (H-3\ell)^{3/2}$. The two parameters
$\ln(\tau_0)$ and $\alpha$ have been obtained from a least-square
fit of the data in a $\ln(\tau)$ versus $(H-3\ell)^{3/2}$ plot.}
\label{fig11}
\end{figure}
Studying the influence of the temperature on $\tau$, we have shown
\cite{Restagno2000} that this time $\tau$ obeys an Arrhenius law
$\tau=\tau_0\exp(\Delta
 \Omega^{\dag}/k_BT)$, where $\Delta \Omega^{\dag}$ is
 identified as the energy barrier for nucleation. In the 2D case,
 we have calculated \cite{Restagno2000} the energy barrier expected from the classical theory of capillarity:
\begin{equation}\label{barr2D}
\Delta
\Omega^{\dag}=\frac{4}{3}\left(\Delta\mu\Delta\rho\gamma_{LV}\right)^{1/2}H^{3/2}
\end{equation}
The $H$ dependence ($\Delta\mu$ being fixed) on the activation
time
 is plotted in fig. \ref{fig11}. As seen in fig.
\ref{fig11}, the DFT model provides  a good agreement with this
classical prediction as far as the distances between the walls $H$
is replaced by an effective distance $H-3\ell$ to take into
account the presence of the wetting films. The prefactor $\alpha$
can be estimated from the data plotted in fig. \ref{fig11},
yielding $\alpha=0.68$, while the classical prediction gives
$\alpha=1.03$ (here the liquid-vapor surface tension -at finite
temperature $T=0.06$- has been computed from independent
Monte-Carlo simulations of the model, yielding $\gamma_{LV}=0.8$).
The macroscopic theory thus gives a correct qualitative picture
and a semi-quantitative agreement of the activation energy for
capillary condensation.

\subsection{Nucleation between parallel plates in three dimensions}
\begin{figure}[htbp]
\centering
\includegraphics[width=7cm]{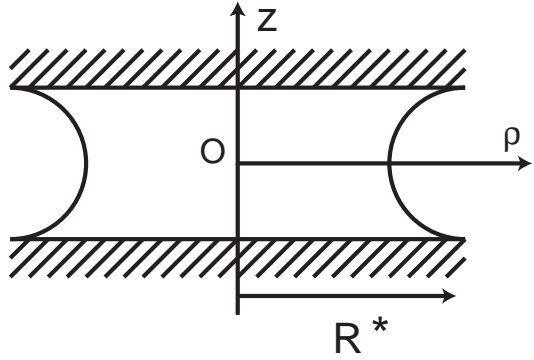}
\caption{Picture of the critical nucleus for capillary
condensation in three dimensions. $R^*$ represent the lateral
extension of the critical nucleus (see text for details). The
total curvature $\kappa$ of the meniscus is equal to
$\kappa=1/R_c=2/H_c$. Note that in 3D, $\kappa$ is the sum of the
in-plane and ``axisymmetric'' curvature.} \label{fig1}
\end{figure}
 We use then the classical theory of capillary  to estimate the activation
 energy for nucleating a liquid phase between two parallel plates. In the grand-canonical ensemble the critical nucleus corresponds
to
 a saddle-point of the grand potential.
 We will consider the perfect wetting situation
 $\gamma_{SV}=\gamma_{SL}+\gamma_{LV}$, although a generalization to the
 partial wetting case is straightforward.
 The grand potential of a pore partially filled with
 liquid may be written \cite{Evans85}
\begin{equation}
\Omega=-p_V V_V-p_L V_L+\gamma_{SV} A_{SV}+ \gamma_{SL}
A_{SL}+\gamma_{LV}
 A_{LV}\end{equation}
 where $V_V$ (resp. $V_L$) is the volume of the gas (resp. liquid) phase and
 $A_{SL}$, $A_{SV}$ and $A_{LV}$ respectively denote the total solid-liquid,
 solid-vapor and liquid-vapor surface area.
  The following expression is
 obtained for the ``excess'' grand potential,  $\Delta\Omega_{tot}=
 \Omega-\Omega_V$, with $\Omega_V$ the grand-potential of the system filled
 with the gas phase only :
 \begin{equation}
 \Delta\Omega_{tot}= \gamma_{LV} A_{LV} + \gamma_{LV} A_{SL}
+\Delta\mu\Delta\rho V_L
 \label{omega2}
 \end{equation}
 where we have used $p_V-p_L\simeq \Delta\rho\Delta\mu$. Within classical capillarity, long range fluid-fluid
 interactions are not taken into account, therefore the critical
 nucleus has to be a liquid bridge connecting the plates.
 One also expects this critical nucleus to exhibit rotational invariance, so that
 $\Delta\Omega_{tot}$ in eq. (\ref{omega2}) is best parameterized in cylindrical
 coordinates (see fig. \ref{fig1}).
 In terms of $\rho(z)$, the position of the LV interface, one obtains
 \begin{multline} \label{Seq1}
 \Delta\Omega_{tot}=  \Delta\rho\Delta\mu 2\pi \int_0^{H\over2}
 dz \rho^2(z)
+ 2\gamma_{LV} \pi \rho^2({H\over2})
 \\+ 2\pi \gamma_{LV} \int_0^{H\over2} dz~\rho(z)\sqrt{1+{\rho_z}^2}
 \end{multline}
 where the index $z$ denotes differentiation.
 Extremalization of the grand potential (\ref{Seq1})
 leads to the usual condition of {\it mechanical equilibrium}, the Laplace
 equation, which relates the local curvature $\kappa$ to the pressure drop according
 to the Laplace law of capillarity:
 ${\gamma_{LV} \kappa}=\Delta p\simeq \Delta \mu\Delta \rho$.
 This condition
 remains valid although the nucleus corresponds to a saddle point and not a minimum of the
 grand-potential.

\begin{figure}[htbp]
\centering
\includegraphics[width=7cm]{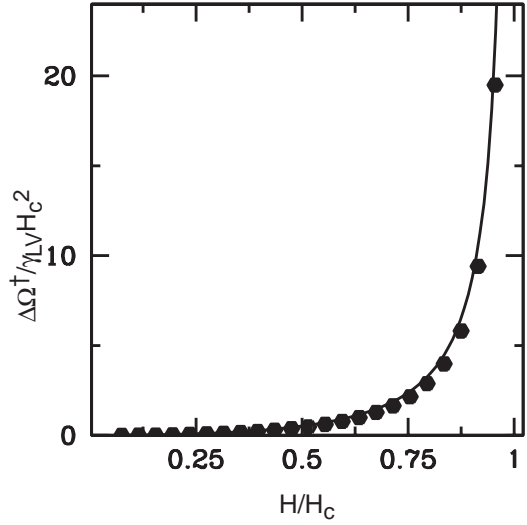}
\caption{Free energy barrier (in 3D) as a function of the
normalized width of the pore, $H/H_c$. The solid line is computed
by numerical integration of the Laplace equation. The points are
obtained from the analytical expression, eq.
(\ref{NRJbarrier3D}).} \label{fig_omega3d}
\end{figure}

 The main difference with bulk homogeneous nucleation comes from the pressure drop
 at the
 interface: here, the liquid pressure inside the meniscus is lower than the gas
 pressure since $\mu<\mu_{sat}$, so that the critical
 nucleus takes the form of a liquid bridge between the solid substrates
 instead of the spherical shape encountered in bulk homogeneous nucleation.
 The previous Laplace equation is non-linear and cannot be
 solved analytically. From dimensional arguments however, one expects
 $\Delta\Omega_{tot}=\gamma_{LV} H_c^2 f(H/H_c)$, with $f(x)$ a
 dimensionless function. The latter can be obtained from the numerical
 resolution of the Laplace equation, yielding the shape of the meniscus
 \cite{numerical}. Numerical integration of eq. (\ref{Seq1})
 then gives the corresponding free energy barrier. The result for the energy
 barrier $\Delta\Omega^{\dag}$ is plotted in
figure \ref{fig_omega3d}. As can be seen from the figure, a
divergence of
 $\Delta\Omega^{\dag}$ is obtained as the pore width $H$ reaches $H_c$. When the axisymetric extension
 of the bridge $R^*=\rho({H\over2})$ is
large compared to $H$, the negative (axisymmetric) contribution to
the curvature is negligible and the L-V profile can be
approximated by a semi-circular shape. This allows to obtain
explicit expressions for the different contributions to
$\Delta\Omega_{tot}$ in eq. (\ref{omega2}) as a function of the
extension $R^*$ of the bridge, namely $V_L=\pi R^{*2}H-{\pi^2\over
4} R^* H^2 +{\pi\over 6}H^3$, $A_{SL}= 2\pi R^{*2}$ and
$A_{LV}=\pi^2 R^*H-\pi {H^2}$. Maximization of
$\Delta\Omega_{tot}$ as a function of $R^*$ yields the following
expression for the free energy barrier
\begin{equation}
\Delta \Omega^\dag = \gamma _{LV} H^2 \biggl[ {\pi^3\over 8}
{\left(1- {H\over{2H_c}}\right)^2 \over {1-{H\over{H_c}} }
}-\left(-{\pi\over3} {H\over{H_c}} +\pi \right)  \biggr]
 \label{NRJbarrier3D}
 \end{equation}
which does exhibit a divergence at $H\sim
H_c=2\gamma_{LV}/\Delta\rho\Delta\mu$. As shown in figure
\ref{fig_omega3d}, this approximate expression is in very good
agreement with the numerical calculation, even at small
confinement $H$. Physically, an important consequence of the
diverging energy barrier at $H_c$ is that the gas phase becomes
extremely metastable: for water at $25^{\textrm o}$C, at a
relative humidity of $p_{vap}/p_{sat}=40\%$, we obtain
$H_c\simeq2$~nm and $\gamma_{LV}H_c^2\simeq70 k_BT$. This
numerical estimate shows that the energy barrier is always larger
than the thermal energy of the system, except when $H/H_c\ll 1$.

\subsection{Experimental evidence of the metastability}
Metastability effects in capillary condensation are very often
observed in porous media, where they are responsible for
hysteresis loops in adsorption/desorption isotherms. In such
hysteresis loops however it is not possible to tell which branch
is the stable one. We have studied the capillary condensation of
alkane vapors between smooth metal surfaces with a Surface Forces
Apparatus \cite{Crassous94}. In this type of experiments, the
pressure of vapor is kept constant, and the distance between the
surfaces (a sphere and a plane) is varied. The condensation of a
liquid phase is revealed by a very strong adhesion force due to
the capillary depression in the liquid phase. On the contrary,
almost no interaction is measured when the surfaces are separated
by the vapor. In these  experiments, one observes the hysteresis
loop of the force as a function of the distance between the
surfaces associated with the metastability effects in the
liquid/vapor transition. The particularity of SFA experiments is
that it is possible to know which branch of the curve correspond
to the lower energy state and therefore which phase is stable for
a given confinement. This is due to the so-called Derjaguin
approximation, which relates the force $F$ measured between a
sphere of large radius of curvature $R$, and a plane, to the
interaction energy per unit area $U$ of two parallel plates
separated by the same distance $H$\cite{Israel}:
\begin{equation}
F(H)/R=4\pi U(H)
\end{equation}
\begin{figure}
\centering
\includegraphics[width=7cm]{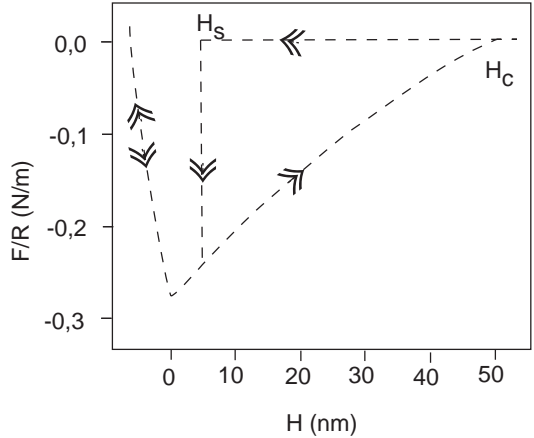}
\caption{Force between a sphere of pyrex of radius $R=3.29$~mm
 covered by a platinum layer and a plane of pyrex in the
presence of a vapor of $n$-heptane.
  The arrows indicate the direction of the surfaces. $H_c$ is the critical distance at
  which there never is a meniscus and $H_s$ the distance at which a  liquid bridge appears when the surfaces are brought into contact.}\label{fig7}
\end{figure}
The force curve shows then clearly that for a large range of
values of the confinement $H<H_c$ the vapor phase is metastable.
In the hysteresis range, the vapor never condenses over the time
of the experiments--typically some hours--showing that the energy
barrier to overcome in order to form a liquid bridge between the
surfaces is actually quite large.
\section{Slow kinetics of capillary condensation: the aging of the friction coefficient}

\subsection{Humidity induced aging of the avalanche angle of a granular medium}
Since the condensation of liquid in a confined geometry can be
hindered by high activation energy, one expects that capillary
condensation processes may display slow kinetics. We discuss here
in more detail the influence of humidity an the slow evolution in
time of a macroscopic property: the avalanche angle of granular
media.
\begin{figure}
\centering
\includegraphics[width=7cm]{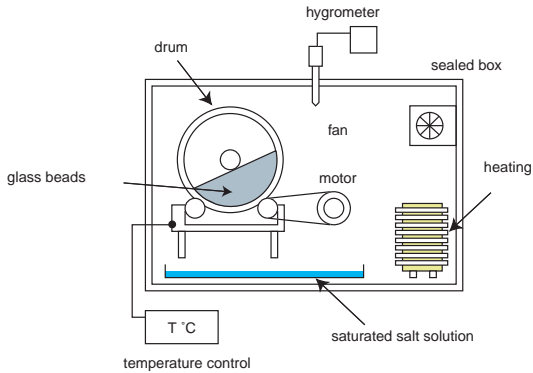}
\caption{Experimental setup. The granular material is placed in a
cylindrical drum with an inner diameter and a length of 10~cm. The
lateral faces of the cylinder are made of glass with an opening at
the center to allow an exchange with the outer atmosphere. The
drum is half filled with glass beads, whose diameter ranges
between 125 and 160 microns.
 The relative humidity $H=p_{vap}/p_{sat}$ is kept at a constant value
 by the method of the saturated aqueous solutions of inorganic salts \cite{cretinon}.} \label{fig4}
\end{figure}

Solid friction properties of a large number of solid materials are
well described by the Coulombs'law, which states that the minimum
tangential force $T$ that must be applied to a contact between
solids in order to bring them into relative motion obeys the
relation $T=\mu_s N$, where $N$ is the normal load applied to the
contact and $\mu_s$ the static friction coefficient. Systematic
studies on various material (for a review see Baumberger
\cite{Baumberger97}) show that $\mu_s$ increases slowly with the
time during which the solid have been into contact at rest. This
aging behavior of the static friction coefficient is often found
to be logarithmic in time
\cite{Rabinowicz}:$\mu_s=b+\alpha\log(t_w)$

More recently, the influence of humidity on this aging behavior
has been brought into evidence. Dieterich \cite{Dieterich84} first
showed in rock onto rock friction , that aging occurs only in
humid atmosphere. In our group, we have studied more specifically
the influence of humidity on aging of friction properties in
granular media. For that purpose, we have measured the avalanche
angle of a granular media consisting of micrometric glass beads in
a controlled environment (see fig. \ref{fig4}). Using a rotating
drum, the time $t_w$ during which the granular heap remains at
rest can easily be varied.

\begin{figure}
\centering
\includegraphics[width=7cm]{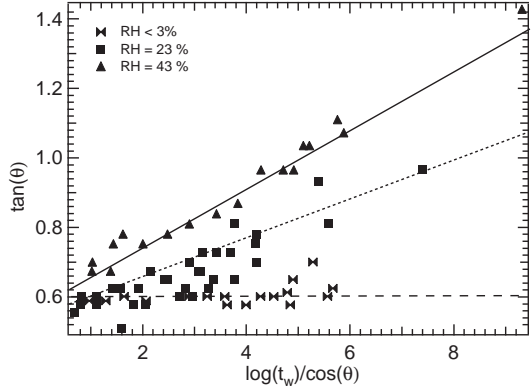}
\caption{The tangent of the maximum stability angle as a function
of
  where $t_w$ is the resting time of the pile of glass beads. The lines are least-square fits
  of the experimental data, whose slope is identified with $\alpha$. Different humidity are plotted
   on the same figure : $H = 3\%$, $H = 22\%$ and  $H = 43\%$.}\label{fig5}
\end{figure}

The evolutions of the maximum angle of stability  as a function of
the resting time $t_w$, at different humidities are shown in
figure \ref{fig5}. The data clearly show an aging behavior, since
the angle of first avalanche systematically increases with the
resting time.

A first important point in this aging behavior is that it cannot
be explained by an increase of the friction coefficient alone.
Indeed, avalanche angle larger than $90^{\textrm o}$ can be
obtained for high humidities and large waiting time; in those
cases the granular heap has enough internal cohesion to remain
stuck to the upper part of internal wall of the drum . The effect
of such a cohesive force can be included in Coulomb's analysis by
adding an additional force $F_c$ to the normal component of the
weight of a layer of glass beads at the surface of the heap
\cite{Halsey}. This yields the following condition for an
avalanche to occur:
\begin{equation}\label{coulomb2}
\tan\theta=\mu_s\left(1+\frac{F_c}{mg\cos\theta}\right)
\end{equation}
where $mg$ is the weight of the layer of glass beads undergoing
the avalanche. Since the data show a linear dependence of the
tangent of the avalanche angle with  $\log(t_w)/\cos\theta$, we
can conclude that there is a cohesion force in the granular media
which increases as $F_c\propto \log t_w$. This dependency is
observed over more than 4 decades of time (resting times range
from 5 to 200 000 s). We do not observe any saturation of the
angle at times as long as 200 000 s. The amplitude of this aging
in characterized by the slope of the lines obtained is this
representation.

\begin{figure}
\centering
\includegraphics[width=7cm]{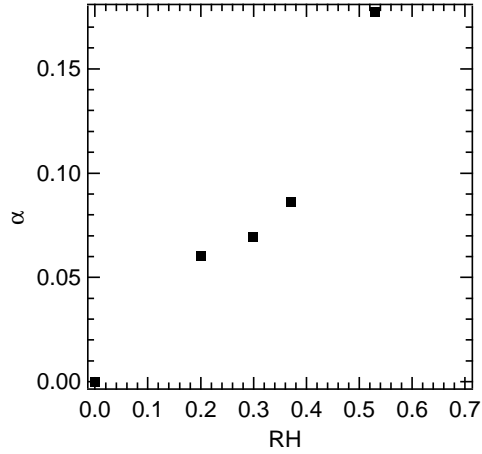}
\caption{Variation of the slope $\alpha(H)$ characterizing the
aging behavior of the first avalanche angle with the relative
humidity pvap/psat.}\label{fig6}
\end{figure}

A very important parameter for this aging behavior is the humidity
of the surrounding atmosphere. We do not observe any aging in dry
atmosphere. Figure \ref{fig6} shows that this aging increases with
the relative humidity: this increase is reversible.

\subsection{A simple model of thermally activated process for aging}
Although a lot can be learned from the perfectly flat slab
geometry, the latter is certainly too idealized to account for the
kinetics of adsorption in ``real'' experimental systems such as
the previous ones. 9
\begin{figure}[htbp]
\centering
\includegraphics[width=7cm]{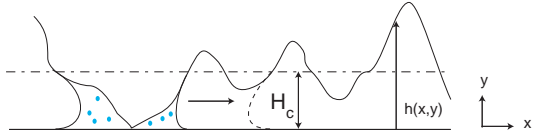}
\caption{Schematic representation of a contact between rough
surfaces. The fact that we consider
  a rough surface on a flat surface does not restrict the generality of this approach.}\label{fig8}
\end{figure}

  We show here that these logarithmic time dependence of the cohesion force between the surfaces
  may be understood by taking into account the influence of surface roughness on the dynamics of capillary condensation. Let us consider a simple model consisting of two surfaces facing each other
and rough  at the nanometric scale, as depicted on fig.
\ref{fig8}. As emphasized in the introduction capillary
condensation typically occurs in pores of nanometric size. We thus
have to  consider the roughness of the surfaces at the {\it
nanometer level}. Here again we shall stay at a macroscopic
description, and focus on a qualitative picture of the influence
of roughness on the transition mechanism. Without loss of
generality, one may consider that one of the walls is perfectly
flat. When roughness is present, there is a broad range of gaps
between the surfaces. In particular, there are regions where the
two surfaces are in close contact. In such regions, condensation
should take place on a very short time-scale. Thus at ``early
times'', one has to consider a set of wetted islands, which we
shall consider as independent one of the other. Once these islands
have formed, they should grow up to a point where the distance
between the surfaces is equal to $H_c$, so that a meniscus of
radius $R_c=H_c/2\cos\theta$ forms at the liquid-vapor interface,
allowing for mechanical equilibrium.

\begin{figure}[htbp]
\centering
\includegraphics[width=7cm]{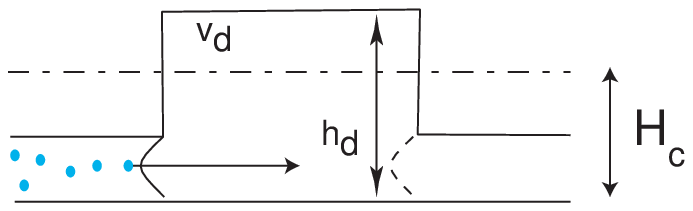}
\caption{Schematic representation of a defect. $v_d$ is the volume
of the region where the height $h_d>H_c$.}\label{fig9}
\end{figure}

In doing so however, the wetted area has to overcome unfavorable
regions where the distance between the two surfaces is larger then
$H_c$. Let us consider a specific jump over such a ``defect'', as
idealized in fig. \ref{fig9}. We denote by $h_d$ the ``averaged''
gap inside the defect ($h_d>H_c$), and by $a_d$ its area. The free
energy cost for the liquid bridge to overcome this defect is
approximatively given by
\begin{eqnarray}
&\Delta \Omega^{\dag} &\simeq a_d \left(\Delta\mu\Delta\rho~e_d-2(\gamma_{SV}-\gamma_{SL})\right)\nonumber\\
& & \equiv v_d \Delta\mu\Delta\rho \label{CCR1}
\end{eqnarray}
where $v_d$ is the excess volume of the defect,
$v_d=a_d~(e_d-H_c)$. We can thus estimate the time to overcome the
defect as
\begin{equation}
\tau=\tau_0 \exp \left\{ {\Delta \Omega^{\dag} \over
{k_BT}}\right\} \label{CCR2}
\end{equation}

One may expect the defects  to have a broad distribution of excess
volume  $v_d$, so that the activation times $\tau$ are very widely
distributed.  After a time $t_w$, only the defects with an
activation time $\tau$ smaller than $t$ have been filled by the
liquid phase. Using eq. \ref{CCR1} and \ref{CCR2} these have an
excess volume $v_d$ which verifies
$v_d<v_{dmax}(t_w)=k_BT(\Delta\mu\Delta\rho)^{-1}\ln(t_w/\tau_0)$.
The number of filled defects at a time $t$ is then typically
$N(t)=v_{dmax}(t_w)/v_0$ where $v_0$ is the typical width of the
distribution of excess volume of the defects. Now, once a liquid
bridge has bypassed a defect, the area wetted by the liquid
increases by some typical (roughness dependent) $a_d$. The time
wetted area grows in time as:
\begin{eqnarray}
&A(t_w) &\simeq N(t_w)a_d  \nonumber \\
&&\frac{a_d}{\Delta\mu/(k_BT)\Delta\rho
v_0}\ln\left(\frac{t_w}{\tau_0}\right) \label{equ10}
\end{eqnarray}
Assuming a typical radius of curvature $\lambda$ of the
asperities, we find if $\lambda\gg r_K$, that $a_d\sim\lambda
r_K$. If we now assume that the vapor is an ideal gas, then
$\Delta\mu=k_BT\ln(p_{sat}/p_{vap})$. This give a cohesion force
between two beads:
\begin{eqnarray}
&F_c(t_w)&=\frac{\gamma_{LV}}{r_K}A(t_w)\nonumber\\
&&\frac{\gamma_{LV}\lambda}{\Delta\rho v_0
\ln(1/H)}\ln\left(t_w/\tau_0\right)
\end{eqnarray}
This model gives the good dependency of the cohesion force with
the resting time $t_w$ in qualitative agreement with the
experiments. It is not possible to test directly this prediction
since $v_0$ and $\lambda$ are not easily measured but we can check
the dependency of $\alpha$ with the relative humidity. However we
can check the internal coherence of this mechanism by estimating
the order of magnitude of the cohesion force needed to induce
aging of the avalanche angle. Using equation \ref{coulomb2}, the
increase of the cohesion force during the time $t_w$ must be:
\begin{equation}
\Delta F_c=\frac{mg\cos\theta}{\mu_s}\Delta\tan\theta(t_w)
\end{equation}
The numerical value of $Delta F_c$ is thus of the order of the
weight of a bead. On the other hand, the maximum value of the
capillary force between two beads in contact is obtained when the
beads are ideal sphere (no roughness):
$F_c=2\pi\gamma_{LV}R=3.4\times10^{-5}$~N for beads of radius
$R=75~\mu$m in water vapor humidity. This is four order of
magnitude larger than the weight of a bead:$4.4\times10^{-9}$N.

Therefore one see that a slow kinetics of capillary condensation
between rough surfaces, leads to a logarithmic growth in time of
the capillary force and can actually be responsible for the aging
of the avalanche angle of granular media, as well as more
generally for the humidity induce aging of the friction properties
of surfaces.

\end{document}